\documentstyle[graphicx]{mn2e}

\newif\ifAMStwofonts
\AMStwofontstrue

\def\pg{{PG1211+143}}
\def\pds{{PDS 456}}

\def\le{{L_{\rm Edd}}}

\def\rg{{R_{\rm g}}}

\def\xmm{{\it XMM-Newton}}

\def\suzaku{{\it Suzaku}}
\def\et{{et al.\ }}

\newcommand{\ls}{\mathrel{\hbox{\rlap{\hbox{\lower4pt\hbox{$\sim$}}}\hbox{$<$}}}}
\newcommand{\gs}{\mathrel{\hbox{\rlap{\hbox{\lower4pt\hbox{$\sim$}}}\hbox{$>$}}}}

\def\Msun{\hbox{$\rm ~M_{\odot}$}}

\def\H0{{\rm ~km~s^{-1}~Mpc^{-1}}}

\def\et{{et al.}}

\title[Detection of an ultra-fast inflow]
      {An ultra-fast inflow in the luminous Seyfert \pg}
\author[K.A.Pounds \et]
       {K.A.Pounds$^{1}$, C.J.Nixon$^{1}$, A.Lobban$^{2}$ \& A.R.King$^{1,3,4}$ \\
$^1$ Department of Physics and Astronomy, University of Leicester, Leicester, LE1 7RH, UK \\
$^2$ Astrophysics Group, School of Physical and Geographical Sciences, Keele University, Keele, Staffordshire ST5 5BG, UK \\  
$^3$ Anton Pannekoek Institute, University of Amsterdam, Science Park 904, 1098 XH Amsterdam, Netherland \\
$^4$ Leiden Observatory, Leiden University, Niels Bohrweg 2, NL-2333 CA Leiden, Netherlands \\}

\date{Accepted ; Submitted }
\pagerange{\pageref{firstpage}--\pageref{lastpage}}
\pubyear{2018}
\begin{document}
\maketitle
\label{firstpage}

\begin{abstract}
Blueshifted absorption lines in the X-ray spectra of AGN  show that ultra-fast outflows with typical velocities $v \sim 0.1c$  are
a common feature of these luminous objects. Such powerful AGN winds offer an explanation of
the observed $M-\sigma$ relation linking the mass of the supermassive black hole and the velocity dispersion in the galaxy’s stellar bulge. An extended \xmm\
study of the luminous Seyfert galaxy \pg\ recently revealed a variable multi-velocity wind. Here we report the detection of a short-lived, ultrafast inflow  during the same
observation. Previous reports of inflows used single absorption lines with uncertain identifications, but this new result identifies an array of resonance absorption lines of highly ionised Fe, Ca, Ar, S
and Si, sharing a common redshift when compared with a grid of realistic photoionization spectra. The redshifted absorption arises in a column of highly ionized matter close to the black hole, with a
line-of-sight velocity, $v \sim 0.3c$, inconsistent with the standard picture of a plane circular accretion disc. This may represent the first direct evidence for chaotic accretion in AGN, where accretion
discs are generally misaligned to the black hole spin. For sufficient inclinations, the  Lense-Thirring effect can break the discs into discrete rings, which then precess, collide and shock, causing near
free-fall of gas towards the black hole. The observed accretion rate for the reported infall is comparable to the hard X-ray luminosity in \pg\, suggesting that direct infall may be a significant contributor
to inner disc accretion.
\end{abstract}

\begin{keywords}
galaxies:active -- galaxies:Seyfert;quasars:general --
  galaxies:individual:PG1211+143 -- X-ray:galaxies
\end{keywords}

\section{Introduction}
It is now well-established that a supermassive black hole (SMBH) lies in the centre of most galaxies, and further that it accretes material through a disc. Over the past 15 years,
observations with a new generation of X-ray Observatories (Jansen 2001, Mitsuda:2007) have revealed ultra-fast outflows (UFOs),  probably
launched from regions of the disc accreting at super--Eddington rates (King \& Pounds 2003) where the momentum in the radiation field released by
accretion can overcome the inward pull from the black hole's gravity. UFOs appear to be a common component of luminous
AGN (Tombesi \et\ 2010, 2011, Gofford \et\ 2013). With typical velocities of $v\sim 0.1c$, these highly ionized winds imply significant feedback onto the
surrounding interstellar gas, offering a likely explanation of the $M-\sigma$ relation (Ferrarese \et\ 2000, Gebhardt \et\ 2000), by simultaneously constraining
the growth of a supermassive black hole and star formation in the central bulge of its host galaxy (King 2003,2005; King \& Pounds 2015).

The archetypal UFO is found in the luminous Seyfert galaxy PG1211+143 (Pounds \et\ 2003). To further explore the properties of this powerful UFO,
the {\it XMM-Newton} X-ray Observatory carried out seven full-orbit ($\sim 100$\,ks) observations over 5 weeks in 2014, with a total on-target exposure
of $\sim$ 630 ks. Full details of observing times,
data reduction procedures and count rates are given in Lobban \et\ (2016). The stacked data revealed a surprisingly complex spectrum,
with the hard X-ray pn camera (Strueder \et\ 2001) finding multiple blue-shifted absorption lines, identified with highly ionized
Fe between $\sim$ 6.6keV and 8.8 keV and outflow velocities of $\sim 0.06c$, $\sim 0.13c$ and $\sim$
0.18c (Pounds \et\ 2016a). Independent support for the multiple outflow velocities was found in higher resolution spectra (Pounds \et\ 2016b) using soft-X-ray data from the co-aligned Reflection Grating
Spectrometer (RGS :den-Herder \et\ 2001). While all previous UFO detections report a single velocity, although with repeated observations sometimes finding a different value, the 2014 \xmm\
observation of \pg\ was
inconsistent with the unique and stable outflow expected from a static axisymmetric accretion
disc (Shakura and Sunyaev 1973, Pounds \et\ 2017). Furthermore, a recent orbit-by-orbit study of the RGS data (Reeves \et\ 2018) has shown significant
inter-orbit variability in outflow column densities over the 5-week campaign, perhaps providing a further indication of short-term inner disc variability.

The highly ionized state of such ultra-fast AGN winds limits strong absorption to the heavier metals, where features of Fe stand out because of its high
astrophysical abundance. For that reason, all current UFO discoveries essentially rest on the detection of
blue-shifted Lyman-$\alpha$ and/or He-$\alpha$ resonance absorption lines  of highly ionized Fe (respective rest energies of 6.70 and 6.96
keV). While spectral modelling, typically over the 2--10 keV band, also includes absorption in lighter metals such as Ca, S, and
Si (eg. Pounds and Page 2006), little attention has been given to other spectral features below $\sim 6$\,keV, where red-shifted Fe K absorption lines
might be seen. The few historical exceptions (Nandra \et\ 1999, Dadina \et\ 2005, Reeves \et\ 2005, Cappi \et\ 2009, Giustini \et\ 2017) are of isolated absorption lines where the
identification -- and hence velocity -- of the absorber remained unclear. In the most detailed report (Dadina \et\ 2005) two (of five)
Beppo-Sax observations of the Seyfert 1 galaxy Mkn 509 detected an absorption line at $\sim 5.5$\,keV (rest-frame), suggesting the feature was
variable on timescales as short as $\sim 20$\,ks, which the authors argued was more easily reconciled with inflowing matter than with a pure
gravitational redshift or failed jet. Of particular relevance to the present study are unidentified absorption lines at $\sim 4.56$\,keV and $\sim
5.33$\,keV in the {\it Chandra} observation of PG1211+143 (Reeves \et\ 2005). While these earlier reports hint at fast-evolving and complicated dynamics in
the inner disc, so far none has provided compelling evidence for a high velocity inflow that could represent a direct challenge to the standard picture of
a circular, planar disc slowly accreting on to the central black hole.

However, recent theoretical arguments suggest that such infall may be common in
AGN, where allowing for the possibility that the accretion disc may be misaligned to the central black hole spin reveals new physical effects. 
Numerical simulations of misaligned discs around a spinning black hole have shown that the Lense-Thirring effect can overwhelm the internal communication of angular
momentum within the disc. This causes the disc to break, leading to individual rings of gas which precess effectively independently -- called disc
tearing (Nixon \et\ 2012, Nixon 2015, Dogan \et\ 2015, Nealon \et\ 2015, Dogan \et\ 2018). When these rings interact, their opposed velocity fields create
shocks which rob the gas of rotational support, allowing it to fall inwards towards the black hole where residual angular momentum causes the gas to
re-circularise at a smaller radius.

Motivated by these theoretical ideas and the paucity of relevant observations, we have carried out a thorough search for rapidly inflowing matter, on a range of
timescales, as part of an on-going orbit-by-orbit analysis of the hard X-ray data from the 2014 {\it XMM-Newton} observation of PG1211+143.

\section {Redshifted absorption} 

The new analysis is based primarily on the high throughput pn camera, while checking for consistency - as necessary - with simultaneous spectral data
from the MOS camera (Turner \et\ 2001) and higher
resolution spectra from the RGS. 
 
As in the published outflow studies, spectral modelling uses the {\tt XSPEC v12} software package (Arnaud 1996), with all spectral fits including absorption  due to the
line-of-sight Galactic column of
N$_{H}=2.85\times10^{20}\rm {cm}^{-2}$  (Murphy \et\ 1996). We assume the reverberation black hole mass estimate of
$4\times 10^{7}$\Msun\ (Kaspi \et\ 2000), and a galaxy redshift of 0.0809 (Marziani \et\ 1996).

The starting point
in modelling each pn data set is the stacked X-ray spectrum (Pounds \et\ 2016a), where the continuum from 1--10 keV is modelled by
a double power law, with discrete spectral features imposed by overlying matter being compared with grids of pre-computed photoionized  absorption and emission spectra, based on
the {\tt XSTAR code} (Kallman \et\ 1996). To facilitate comparison with previous outflow analyses (eg Pounds \et\ 2016a), for the pn and MOS data we use the publically available grid25 (v21),
which assumes a power law continuum of $\Gamma$ = 2 and a fixed turbulence velocity 200 km s$^{-1}$, and custom-built grids derived from the observed source spectrum (Pounds \et\ 2016b) for the RGS
spectral analysis. Free parameters in the spectral fitting include the ionization, column density and observed redshift (or blueshift) of photoionized matter, with continuum normalisation also free to
vary on re-fitting. 

We find no significant evidence for strongly 
redshifted absorption \footnote {Potential confusion with strong Fe K emission constrains the resolution of Fe K absorption lines with present data, setting a lower redshift limit
$\sim$ 0.15}
in the first observing period (spacecraft orbit number 2652, or rev2652), but that outcome changed for rev2659, just 
two weeks later.

\subsection{A strong inflow in rev2659}

Modelling the 1--10 keV pn spectrum from rev2659 revealed a strong absorption component at the extreme redshift $\sim$0.483$\pm$0.008. With a high column density (N$_{H} \sim 4.3\times10^{23}\rm cm^{-2}$)
and high ionization parameter (log $\xi \sim 3.7$ erg cm s$^{-1}$), the red-shifted absorber is highly significant, its inclusion improving the fit by $\Delta
\chi^{2}$ of 19/3. The corresponding probability of a false detection (P=4.4 $\times 10^{-4}$) is a physically robust measure, being based on a comparison of the observed broad band absorption
with a set of physically realistic model photoionization spectra.

To check that the model was not impaired by some unknown data artefacts, the redshifted
component was removed from the {\tt XSTAR} model and the ratio of data-to-model visually examined. 
Figure 1 (top) reproduces that ratio plot, showing an array of absorption lines identified with resonance transitions in H- and He-like ions of Fe,  Ca,
Ar, S and Si. The two strongest absorption lines are identified with Lyman-$\alpha$ and He-$\alpha$ of Fe, with individual redshifts closely matching the {\tt XSTAR} model component, while
the array as a whole yields a weighted mean observed redshift of 0.476$\pm$0.005.  The Ca line at $\sim$2.6 keV is most likely a blend of
H- and He-like resonance lines. Table 1 lists each 
observed line energy, obtained by scanning a narrow negative Gaussian across the ratio plot, with the individual
line widths set comparable to the pn detector resolution at each photon energy. We note the detector resolution corresponds to an intrinsic line width of order 1500 km s$^{-1}$ at 5 keV, substantially
greater than the  preferred {\tt XSTAR} model value, indicating the observed
absorption line widths are essentially unresolved in these data.

\begin{figure}
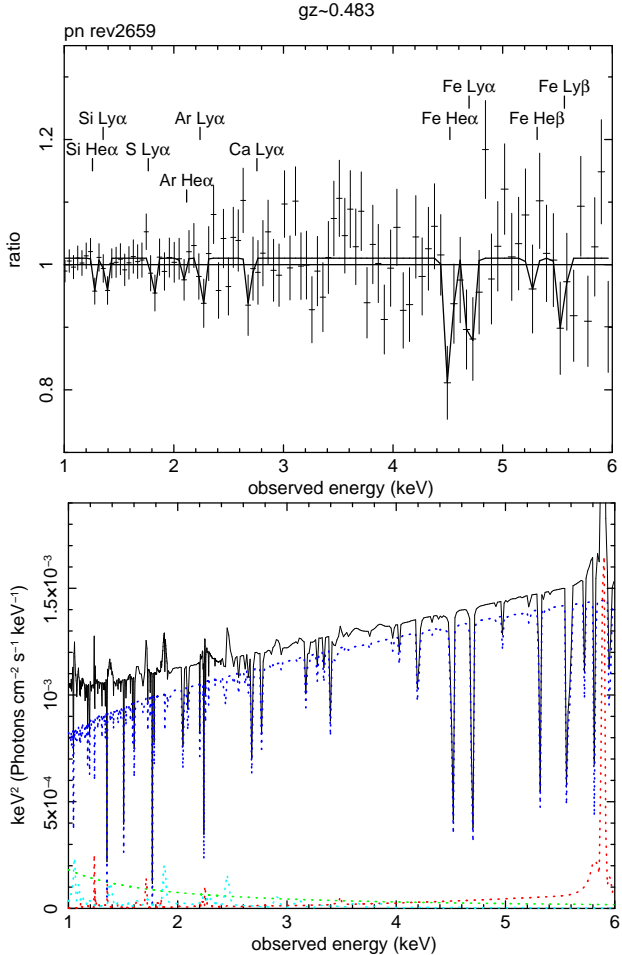
                                  
\centering                                                              
\includegraphics[width=6.7cm, angle=270]{fig1_redshift_ratio.ps}
\centering                                                              
\includegraphics[width=6.03cm, angle=270]{fig1_redshift_model.ps}
\caption{(top)Ratio of pn data-to-model for the rev2659 observation when the absorption component with redshift gz$\sim$0.483 is 
removed and the model re-fitted. Absorption lines identified with resonance K-shell transitions in Fe, Ca, Ar, S and Si yield a weighted mean redshift of 0.476$\pm$0.005.
(lower) The initial spectral model, with a hard power law (dark blue) showing the imprint of the redshifted absorber, the unabsorbed soft power law found in difference spectra (green),
and emission from an ionized reflector (red) and the photoionized outflow (light blue).} 
\end{figure}

\begin{table}
\centering
\caption{Narrow Gaussian lines sequentially fitted to the identified absorption features in the rev2659 pn data shown in Figure 1. Line widths were set comparable to the relevant pn detector
resolution
and all line energies are in keV. The proposed identification and corresponding redshift
for each line are listed, together with the related improvement in $\Delta \chi^{2}$ after re-fitting. The relatively high redshift of the Ca Ly-$\alpha$ line may
be explained by a blend with the corresponding He-$\alpha$ line, though its low statistical weight does not affect the array redshift}
\begin{tabular}{@{}lcccc@{}}
\hline
line i.d. & obs energy  & lab energy  & obs redshift & $\Delta \chi^{2}$ \\
\hline
Fe Ly-$\alpha$ & 4.701$\pm$0.010 & 6.96 & 0.481$\pm$0.004 & 8/2 \\ 
Fe He-$\alpha$ & 4.511$\pm$0.011 & 6.703 & 0.486$\pm$0.004  & 15/2 \\ 
Fe Ly-$\beta$ & 5.543$\pm$0.020  & 8.25 & 0.488$\pm$0.005  &  4/2  \\ 
Fe He-$\beta$ & 5.27$\pm$0.08  & 7.88 & 0.495$\pm$0.023   &  1/2 \\ 
Ca Ly-$\alpha$ & 2.67$\pm$0.025 & 4.09 & 0.532$\pm$0.014 &  2/2 \\
Ca He-$\alpha$ & 2.67$\pm$0.025 & 3.903 & 0.462$\pm$0.014 &  2/2 \\
Ar Ly-$\alpha$ & 2.257$\pm$0.009 & 3.32 & 0.471$\pm$0.006  &  5/2 \\
Ar He-$\alpha$ & 2.09$\pm$0.09 & 3.140 & 0.50$\pm$0.06  &  1/2 \\
S  Ly-$\alpha$ & 1.814$\pm$0.010 & 2.62 & 0.452$\pm$0.016  &  4/2 \\ 
Si Ly-$\alpha$ & 1.382$\pm$0.007 & 2.006 & 0.452$\pm$0.008 &  5/2 \\
Si He-$\alpha$ & 1.283$\pm$0.031 & 1.865 & 0.454$\pm$0.025 &  7/2 \\  
\hline
\end{tabular}
\end{table}

Allowing for the cosmological redshift of \pg\ (0.0809) the observed redshift of $0.48\pm0.01$ corresponds to a Doppler-corrected inflow velocity v$\sim$0.30$\pm$0.01c. As this velocity is
substantially higher than any of the 3 contemporary outflow velocities, we suggest the matter being a 'failed outflow' relatively unlikely, since a lower velocity outflow -launched from a
larger radius - would have too much angular momentum to fall back at the higher velocity. The more
interesting alternative is absorption in matter accreting from out-of-the-plane of the disc, and in line of sight to
the hard X-ray continuum source (conventionally assumed to be a compact hot corona above the inner disc).  

In the context of matter falling freely towards the SMBH, equating the infall velocity to the 
local free-fall velocity places the absorber near 20 $\rg$, where $\rg$ is the gravitational radius.   We note that $\sim5\%$ of the observed redshift would be
gravitational at that location.

Independent support for the above finding is potentially available from the co-aligned MOS camera (Turner \et\ 2000), which was operated throughout the 2014 campaign. An initial check on the rev2659
MOS data did indeed find a highly redshifted absorption component, consistent with that reported from the pn data analysis, although the outflow components are less well constrained due to the lower
sensitivity of the MOS above $\sim$6 keV. To obtain a common spectral fit with both data sets, we therefore re-fitted the pn model, described above, with the addition of the MOS data, and allowing the
joint parameters to vary. The outcome was positive,
with the redshifted absorber increasing in significance ($\Delta\chi^{2}$ of 26/3), while retaining a similar column density (N$_{H}$$\sim$$6.3\times10^{23}\rm cm^{-2}$) and ionization parameter
(log $\xi \sim 3.4$ erg cm s$^{-1}$). The false detection probability for the redshifted absorption component in the joint spectral fit
is now further reduced with P $\leq10^{-5}$.

\subsection{Absorption in the RGS soft x-ray spectrum of rev2659}

Finally, higher resolution soft X-ray spectra (albeit with lower count rates) from the RGS instrument, also co-aligned with the pn camera, are examined to check for possible lower ionization matter
in the infalling stream. We recall that detection of cooler embedded matter provided important confirmation of multiple velocities in the
outflow spectrum of \pg\ (Pounds \et\ 2016b).

Again starting with the mean 2014 outflow spectrum, the rev2659 RGS data was re-modelled, now allowing for an additional
red-shifted absorber.
That search was successful, with a soft X-ray absorption component ($\Delta \chi^{2}$ of 14/3; P=3 $\times 10^{-3}$) being found at the same extreme redshift
($\sim 0.483 \pm 0.003$), though with a much lower column density (N$_{H} \sim 3\times10^{21}\rm cm^{-2}$) and ionization parameter (log $\xi \sim 0.95$ erg cm. s$^{-1}$).
Inclusion of the redshifted absorber improved the overall 12--35 \AA\ soft X-ray spectral fit by $\Delta
\chi^{2}$ of 14/3, perhaps again indicating 
higher density matter
embedded in the primary inflow. Interestingly, a soft X-ray emission component is also marginally significant ($\Delta \chi^{2}$ of 6/2) with the same redshift.

Confidence in the highly red-shifted soft X-ray absorption and emisiion components is supported by the detection of several key features in the spectral residuals, when the high redshift components are
removed from the
{\tt XSTAR} model fit (Figure 2). While the plot is much 'noisier' than for the pn hard X-ray data, the strong 1s-2p resonance absorption lines of He-like O and Ne are both indicated at a redshift
consistent with the {\tt XSTAR} model
value, while the broad absorption line at $\sim$24.2 \AA\ corresponds to a similar redshift when identified with the Fe-M UTA, albeit with lower precision as the rest wavelength of
the UTA depends on the ionization distribution. Narrow emission lines identified with of O Ly-$\alpha$ and the forbidden line of the OVII triplet also match the same inflow redshift.

\begin{figure}                                  
\centering                                                              
\includegraphics[width=6.7cm, angle=270]{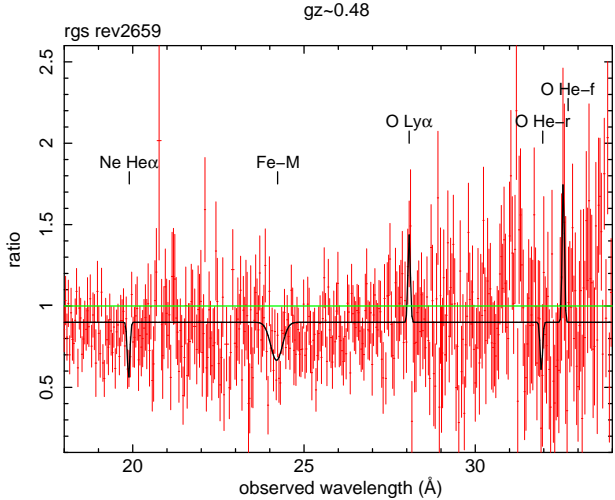}
\caption{Ratio of RGS data-to-model for rev2659 when absorption and emission components with redshift z$\sim$0.48 are
removed. Absorption lines identified with the 1s-2p resonance lines of He-like O and Ne are detected at a redshift consistent with the {\tt XSTAR} model value.
The broad absorption line at $\sim$24.2 \AA\ corresponds to a similar redshift when identified with the Fe-M UTA, although with lower precision as the rest wavelength of
the UTA depends on the ionization distribution. Narrow emission lines of OVIII Ly-$\alpha$ and the forbidden line of the OVII triplet also match the model redshift. The line
markers are located at wavelengths corresponding to a redshift of 0.48. } 
\end{figure}

Two less strongly red-shifted components are also found in the
RGS spectrum, at z$\sim$0.19$\pm$0.01 ($\Delta\chi^{2}$ of 17/3: P=$7\times 10^{-4}$) and z$\sim$0.17$\pm$0.01 ($\Delta\chi^{2}$ of 15/3: P= $1.8\times 10^{-3}$).
While the lower implied velocities might simply represent a line of sight at a larger angle to the flow stream, the required large deviation in the flow vector seems
unlikely so close to the SMBH.

Alternatively, assuming each flow is along the line of
sight, the corresponding inflow
velocities would $\sim$0.080c and $\sim$0.097c. Both components are highly ionized, with log$\xi$ $\sim$3.4 erg cm s$^{-1}$, similar to that for the pn
detection, but with respective column densities of N$_{H} \sim 4.6\times10^{21}\rm cm^{-2}$ and N$_{H} \sim 1.1\times10^{21}\rm cm^{-2}$, much less than in the pn
detection.

An intriguing speculation is that the lower velocity and column density of these soft X-ray inflows represent the primary, highly ionized flow seen
along a different 
sight line (in this case to the soft 
X-ray continuum), providing an upstream view through an accelerating flow 
approaching the SMBH. Support for that conjecture is provided by the absence of a higher energy spectral counterpart to the lower redshift soft X-ray absorbers in the pn data. 

If a true up-stream measure, the lower velocity of the highly ionized soft X-ray absorber, v$\sim$0.10c, would correspond to a radial 
distance for the
absorber of $\sim$200 $R_g$, conceivably on a sight-line to the larger scale, thermal disc
emission.  The much lower column density  might then be qualitatively consistent with a converging - as well as accelerating inflow.
While such deductions are limited by having only two spectral data points, the potential of future observations with higher spectral resolution and photon grasp is
clear.

\section{Evidence of redshifted absorption in later orbits}

The pn data from the third observation, in rev2661, was examined as above, finding only an upper limit for an inflow at a redshift near $\sim$0.48, with (N$_{H} \leq 7\times10^{21}\rm cm^{-2}$).
There was, however, a significant redshifted component at z $\sim$0.36$\pm$0.01, with a similar ionisation parameter (log $\xi\sim 3.5$ erg cm s$^{-1}$, but lower column density
(N$_{H} \sim 1.5\times10^{23}\rm cm^{-2}$) than in rev2659. Adding
this red-shifted component to the {\tt XSTAR} spectral model improved the fit statistic by  $\Delta \chi^{2}$  of 9/3 (Table 2). The physical reality of this ionized absorber was again supported by 
the identification of resonance absorption lines of He- and H-like ions of Fe,  Ca, S and Si, with measured redshifts consistent with the {\tt XSTAR} value.
Assuming a flow close to the line of sight to the hard X-ray corona, the inflow velocity in rev2661 would be
v$\sim 0.23 \pm$ 0.01c. While it is not possible to relate this inflow to the strong component seen in rev2659,
the significant change in both inflow velocity and column density over $\sim$4 days suggests a separate inflow, and fine spatial or temporal
structure in the flow close to the SMBH. 

To complete the search for redshifted absorption, the pn data analysis was repeated for the remaining four \xmm\ orbits, revs2663, 2664, 2666 and 2670,
spaced at intervals of $\sim$4, $\sim$2, $\sim$4 and $\sim$8 days, respectively. Significant absorption was found at z $\sim$0.34$\pm$0.02
in rev 2666, but with a lower ionisation parameter and column density (than rev 2661).  Table 2 summarises the pn inflow detections, with that 
in rev2659 being the most remarkable, in terms of redshift and a large and variable column density, with the simultaneous MOS and RGS data providing independent support for
the $\sim$0.30c inflow.

All redshift (and inflow velocity) measurements are formally lower limits, as they assume the flow is aligned with the line of sight. However, we suggest this is likely to be the case
for inflowing matter close to the black hole and adjacent corona.

\begin{table*}
\centering
\caption{Summary of inflow components detected in modelling of pn spectra from 3 individual spacecraft orbits and 3 sub-sections of the orbit 2659 
data from the 2014 {\it XMM-Newton} campaign. The final two columns show the improvement in $\chi^{2}$ by inclusion of the redshifted absorption component and 
the corresponding false detection probability} 
\begin{tabular}{@{}lcccccc@{}}
\hline
rev     & obs redshift     & N$_{H}\times 10^{23}$ & $\log\xi$      & velocity ($c$) & $\Delta\chi^2$ & P value \\
\hline
2659    & $0.483\pm 0.008$ & $5.6\pm 2.4$       & $3.48\pm 0.05$   & $0.30\pm 0.01$  & 18/3     & $4\times 10{-4}$ \\ 
2659(1) & $0.484\pm 0.014$ & $2.6\pm 3.5$       & $3.4\pm 0.1$   & $0.31\pm 0.01$  & 10/3     & $2\times 10{-2}$ \\ 
2659(2) & $0.485\pm 0.012$ & $5.7\pm 3.6$       & $3.5 (f)$      & $0.31\pm 0.01$  & 6/2      & $5\times 10{-2}$ \\ 
2659(3) & $0.450\pm 0.009$ & $18\pm 5$          & $3.5\pm 0.1$   & $0.29\pm 0.01$  & 20/3     & $2\times 10{-4}$ \\ 
2661    & $0.35\pm 0.01$   & $5.3\pm 2.3$       & $3.50\pm 0.17$ & $0.22\pm 0.01$  & 9/3      & $3\times 10{-2}$ \\ 
2666    & $0.34\pm 0.02$   & $0.3\pm 0.1$       & $2.6\pm 0.2$   & $0.21\pm 0.01$  & 14/3     & $3\times 10{-3}$ \\ 
\hline
\end{tabular}
\end{table*}

\section{Discussion}

While several previous observations have suggested the presence of single redshifted absorption lines in AGN X-ray spectra, we report here the first detection of a physically realistic
redshifted absorption
line spectrum, consistent with highly ionized matter falling inwards at v$\sim$0.3c.  Such an unambiguous detection of strongly red-shifted X-ray absorption in an AGN provides direct
evidence for substantial highly ionized matter close
to - and apparently  converging on - the SMBH. Variability on timescales of hours, as described below, indicates fine spatial or temporal structure in the matter crossing the line
of sight. Here we briefly consider the implication of this result, and why such
clear evidence has remained elusive until now.

The detection of ultra-fast and highly ionized outflows (UFOs) in early X-ray spectra of the luminous, narrow line Seyfert galaxy \pg\ (Pounds \et\ 2003) and
QSO \pds\ (Reeves \et\ 2003), opened a new field of study of AGN, made possible by the uniquely high throughput  of X-ray
spectrometers on ESA's \xmm\ X-ray Observatory, launched in late 1999. King and Pounds (2003, 2015) note that such winds are a natural result of a high accretion rate,
with excess matter being driven off by radiation pressure when the accretion rate exceeds the local Eddington limit (Shakura and Sunyaev
1973).

While this picture provides a satisfactory explanation of most UFOs, with each observation yielding a unique wind velocity, an extended
5-week study of \pg\ in 2014 found a more complex outflow profile, with velocities of $\sim$0.06c, $\sim$0.13c and $\sim$0.18c detected in the
stacked data set. Such complexity was clearly inconsistent with a wind profile launched from the classical (flat axisymmetric) accretion disc of Shakura \& Sunyaev, suggesting some intrinsic disc
instability or rapidly variable accretion rate in the inner disc. As noted above, rapidly varying accretion could be a
direct consequence of the way in which AGN accrete, where matter falls towards the SMBH with essentially random orientations (King and
Pringle 2006, 2007). 

In that case Lense-Thirring precession will cause misaligned orbits to precess around the black hole spin vector, with the inner disc warping and rings of matter breaking free.
Collisions between 
neighbouring rings might then lead to matter falling inwards to a new radius defined by its residual  angular
momentum, where it may form a new disc. Simulations suggest material added to the inner disc from the `disc-tearing' region
can result in a substantial increase in the local accretion rate, allowing an otherwise sub-Eddington flow to become briefly 'super-Eddington'.  In that regard it is interesting to note that the
distribution of UFO velocites from the \xmm\ and \suzaku\ archival searches (King and Pounds 2015; fig.4) is consistent with the expected range of tearing radius.
With precession timescales (for a black hole mass $4\times10^7M_\odot$ and spin parameter 0.5) predicted to be of order weeks for radii of order 50 $\rg$, to
a few years for radii of order  200 $\rg$ (Nixon \et\ 2012), the simultaneous detection of 3 primary wind velocities  in the 2014 \xmm\
observation of \pg\ appears feasible, with the hard X-ray source being viewed through 3 independently expanding shells of ionized gas. 

We suggest the detection of strongly red-shifted X-ray absorption spectra, reported here, may represent the first direct observational evidence of disc tearing and thereby of
the importance of random or 'chaotic accretion' in AGN. 
While the observation is of high significance only in the \xmm\ orbit rev2659, the extreme redshift ($\sim$0.48) and inflow velocity ($\sim$0.3c)
indicates absorption in a significant body of matter plunging towards the SMBH at $\sim$20Rg. Confidence in that detection is increased when the independent MOS data are included in a joint spectral
spectral fit, while further
support is provided by the RGS detection of a smaller column of less ionized (presumably, higher density) matter, co-moving 
with the primary inflow.

The transient nature of the rev2659 inflow largely explains why a compelling detection has not been reported before. Isolated claims are not uncommon, however, with several
single line detections noted in the Introduction, and both \xmm\ and \suzaku\ archival searches finding - but not discussing - transient absorption
lines in the region occupied by
red-shifted Fe K lines. Uniquely, the observation in rev2659 has both a sufficiently large redshift and high column density to reveal multiple
absorption lines that - when matched with resonance lines of Fe, Ca, Ar, S and Si - share a common redshift. In quantifying that spectrum, the {\tt XSTAR} analysis is key,
since only by comparison with a set of physically realistic ionized absorption spectra can the significance be properly assessed. A valuable bonus of such detailed modelling
is in quantifying the key parameters (ionization, column density, apparent redshift) of the flow.

In the context of chaotic accretion, where disc tearing and subsequent shocking is predicted to lead to streams of matter falling towards the SMBH from
essentially random directions, the highly variable nature of redshifted absorption, seen along a sight-line to the centrally located hard X-ray continuum source, is perhaps no
surprise. As each stream plunges towards the black hole the increasing density will allow a greater fraction of the accreting matter 
to pass across the line of sight. While the frequency of detections during the 5-week observation of \pg\ suggests multiple events, the data from rev
2659 offers the best quantitative measure of an individual accreting stream, with the predicted timescales for disc tearing placing this in terms of the overall accretion rate of PG1211+143.

If the conditions for disc tearing are satisfied, the infall of a single gas cloud should be
detectable for a time of order $\sim R/c$, at intervals of $\sim \omega_{LT}^{-1}$ ($\sim 200 r$\,s and $\sim 200 r^3$\,s respectively, for a BH
mass $4\times 10^7M_\odot$, where $r = R/R_g$). For $r\sim 200$, we can expect the  observed inflow to be maintained for much of the 100\,ks rev2659
observation, but not over the 4 day gap to the subsequent observation, in rev2661. To check this prediction we re-examined the rev2659
data in 3 roughly equal intervals, finding that -- while maintaining a consistent redshift -- the column density rose strongly from the first
through to the third time sections, with the final $\sim 30$\,ks being nearly optically thick to Thomson scattering. In contrast, by rev2661 the
$v\sim 0.3c$ inflow was gone.

We use the higher resolution flow data of rev2659 (Table 2) to estimate the peak mass flow rate, where -- as a stream of matter plunges towards the
black hole -- its increasing compactness ensures the third segment represents a maximum fraction passing through the line
of sight. We assume a cylindrical inflow element at a radial distance $20 R_g$, with length $5R_g$ constrained by the well-defined velocity, and
diameter $2 R_g$ to allow a reasonable chance of detection along a line of sight to the hard X-ray source. The mean particle density is then
$N_H/5R_g$, and the observed mass element $m_{\rm in} = ( {\rm
volume}~\times~{\rm density}~\times~{\rm proton~mass}) \sim 15R_g^{3}\times (3.6\times10^{23}/R_g)\times 1.7\times10^{-24}$\,g. For a  black hole
mass of $4\times10^7 M_\odot$, $R_g \sim 6\times10^{12}$\,cm and $m_{in}\sim 3.3 \times10^{26}$\,g. Since the observed element will cross the line of sight in
$\sim 3000$\,s, the instantaneous observed inflow mass rate is $\sim 10^{23}\,{\rm g\,s}^{-1}$.

Assuming an accretion efficiency $\eta\sim 0.1$, the observed accretion rate would yield a luminosity of $\sim 10^{43}\,{\rm
erg\,s^{-1}}$, comparable to the hard X-ray continuum of \pg, the primary inner disc emission component. In
the context of disc tearing each pair of torn-off rings may
interact many times, producing many clouds and a more continuous stream of material, although with only a small fraction likely to fall
through an observer's line of sight. 

Evidence for additional inflowing matter found in two of the remaining 5 {\it XMM-Newton} observations in 2014, with lower
velocities and column densities (than in rev2659) might indicate infalling matter further from the SMBH. Details of all the positive detections are
listed in Table 2, revealing a picture of one remarkably strong -- and rapidly variable -- inflow (in rev2659), which we suggest represents
an infalling matter stream within $\sim 20 R_g$ and lying close to the line of sight, and lower redshift absorption $\sim$4 and $\sim$14 days
later, perhaps representing matter still having too much angular momentum to accrete.

Finally, we note the most likely alternative to the above picture, that of a failed outflow, is unlikely given that the inflow velocity in rev 2659 is much higher than
any observed outflow component in \pg. Assuming that the latter arise from larger disc radii (with corresponding escape velocities), that material would have too much angular momentum to acquire the higher
in-fall velocity observed. For the 'failed-wind' model to explain the data then requires the same velocity outflow to be unobservable for precisely the period in which the inflow is observed; that
seems unlikely, requiring an uncomfortable degree of fine-tuning.

\begin{figure}                                  
\centering                                                              
\includegraphics[width=6.3cm, angle=270]{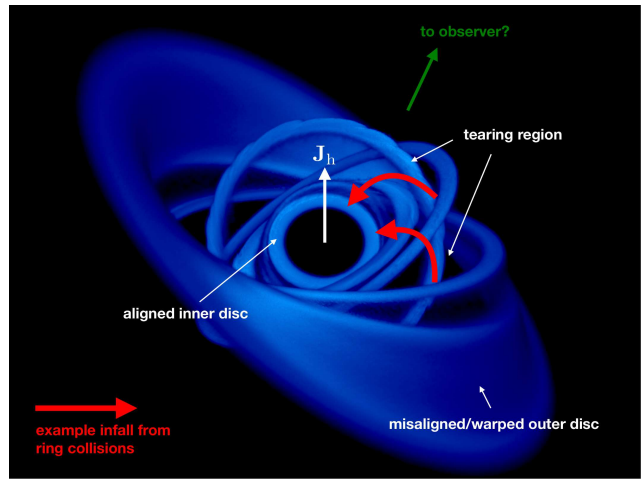}
\caption{Characteristic disc structure from the simulation of a misaligned disc around a spinning black hole. The outermost regions are warped and
  remain misaligned. Inside this, several rings have broken free and are freely precessing through the Lense-Thirring effect. The innermost material
  has fallen from the shocks which occur between rings, and is aligned to the central black hole spin (it is the misaligned component of angular momentum
  which is cancelled in the shocks -- and transferred to the hole through precession). Depending on the observer's line of sight, the infalling matter may or may not
  obscure the central emitting regions. Note that the black hole spin vector is drawn as an arrow up the page, but this is the projection on to the page.
  The black hole spin vector points out towards the reader as well as in the direction drawn, such that the vector is normal to the inner-most ring of gas.} 
\end{figure}

\section{Conclusions}
The detection of strongly red-shifted X-ray absorption in data from an extended \xmm\ observation of the luminous Seyfert galaxy \pg\ has provided compelling evidence
for the inflow of matter at $v \sim 0.3c$. We suggest a possible origin for the inflow in the form of disc tearing, arising when an accretion disc misaligned to the
spin plane of the SMBH precesses due to the Lense-Thirring effect (frame-dragging), with the precession close to the hole being fast enough for individual rings to break off
(Nixon \et\ 2012, Dogan \et\ 2018). Collisions between independently precessing rings then cause shocks and a loss of rotational support and the subsequent infall of disc gas.
If this picture is confirmed, there are strong implications for the fuelling of AGN, and the evolution of SMBH masses and spins. For example, misaligned or `chaotic' accretion
would in general allow the spin to remain low (King \et\ 2008), allowing efficient mass growth for the black hole. This might then remove the need for massive SMBH seeds in the
early Universe (King \& Pringle 2006, 2007).

The transient nature of the strong inflow detected in the 2014 {\it XMM-Newton} data, with variability on a timescale of hours, helps explain why a compelling
detection has not been reported
before. It may also be relatively rare to have a sight-line to the hard X-ray source that passes through infalling matter so close to the black
hole, and hence with such a high particle density, as we found in rev2659.

Looking ahead, X-ray absorption spectra obtained over different sight lines, with high cadence and resolution, together with parallel advances 
in theory and computation, offer the exciting prospect of
mapping the fine structure of ionized flows and accretion close to the SMBH. The high spectral 
resolution of the microcalorimeter planned for the forthcoming Japanese {\it XRISM} mission 
will be well matched for studying the highly ionized matter characteristic of fast inflows. Such new data together with new and a broader search of archival \xmm\ observations should
provide the best opportunity for further exploration of accretion disc physics and black hole growth in AGN prior to the launch of {\it Athena}, a 
decade hence.

\section*{ Acknowledgements }
\xmm\ is a space science mission developed and operated by the European Space Agency and we acknowledge the  outstanding work of ESA staff in
Madrid in successfully planning and conducting the observations which underpin this paper. Our thanks are due to colleague Simon Vaughan for enlightening discussions on how best to present the
statistical aspects of the reported analysis. CN holds an Ernest Rutherford Fellowship 
(ST/M005917/1) and AL is supported by a consolidated grant to Keele, both funded by the UK Science and Technology 
Facilities Council. Finally, we thank the referee for a careful and perceptive review of the initial manuscript.


\begin{thebibliography}{}
\bibitem{} Arnaud K.A. \ 1996, ASP Conf. Series, 101, 17
\bibitem{} Cappi M. \et \ 2009, A\&A, 504, 401
\bibitem{} Dadina M \et \ 2005, A\&A, 442, 461
\bibitem{} Dogan S., Nixon C., King A., Price D.J. \ 2015, MNRAS, 449, 1251
\bibitem{} Dogan S., Nixon C., King A., Pringle J.E \ 2018, MNRAS, 476, 1519
\bibitem{} den Herder J.W. \et  \ 2001,  A\&A, 365, L7
\bibitem{} Ferrarese L. \& Merritt D. \ 2000, ApJ, 539, L9
\bibitem{} Gebhardt K. \et \ 2000, ApJ, 539, L13
\bibitem{} Giustini M. \et \ 2017, A\&A, 597, A66
\bibitem{} Gofford J., Reeves J.N., Tombesi T., Braito V., Turner T.J., Miller L., Cappi M. \ 2013, MNRAS, 430, 60
\bibitem{} Jansen F. \et \ 2001,  A\&A, 365, L1
\bibitem{} Kallman T., Liedahl D., Osterheld A., Goldstein W., Kahn S. \ 1996, ApJ, 465, 994
\bibitem{} Kaspi \et \ 2000, ApJ, 533, 631
\bibitem{} King A.R. \ 2003, ApJ, 596, L27
\bibitem{} King A.R. \ 2005, ApJ, 635, L121
\bibitem{} King A.R. \& Pringle J.E. \ 2006, MNRAS, 373, 90
\bibitem{} King A.R. \& Pringle J.E. \ 2007, MNRAS, 377, 25
\bibitem{} King A.R., Pringle J.E. \& Hoffman J.A. \ 2007, MNRAS, 385, 1621
\bibitem{} King A.R. \& Pounds K.A. \ 2003, MNRAS, 345, 657
\bibitem{} King A.R. \& Pounds K.A. \ 2015, ARA\&A, 53, 115
\bibitem{} Lobban A.P, Vaughan S., Pounds K.A., Reeves J.N. \ 2016, MNRAS, 457, 38
\bibitem{} Marziani P., Sulentic J.W., Dultzin-Hacyan D., Clavani M., Moles M. \ 1996, ApJS, 104, 37
\bibitem{} Mitsuda K. \et \ PASJ, 59, S1
\bibitem{} Murphy E.M., Lockman F.J., Laor A., Elvis M.\ 1996, ApJS, 105, 369
\bibitem{} Nandra K., George I.M., Mushotzky R.F., Turner T.J., Yaqoob T. \ 1999, ApJ, 523, L17
\bibitem{} Nealon R., Price D.J., Nixon C. \ 2015, MNRAS, 448, 1526
\bibitem{} Nixon C. \ 2015, MNRAS, 450, 2459
\bibitem{} Nixon C., King A.R., Price D., Frank J. \ 2012, ApJ, 757,24
\bibitem{} Pounds K.A., Reeves J.N., King A.R., Page K.L., O'Brien P.T., Turner M.J.L. \ 2003, MNRAS, 345, 705
\bibitem{} Pounds K.A. \& Page K.L. \ 2006, MNRAS, 360, 1123 
\bibitem{} Pounds K.A, Lobban A., Reeves J.N., Vaughan S.A. \ 2016a, MNRAS, 457, 2951 
\bibitem{} Pounds K.A, Lobban A., Reeves J.N., Costa M., Vaughan S.A.\ 2016b, MNRAS, 459, 4389
\bibitem{} Pounds K.A., Lobban A.P., Nixon C. \ 2017, Astron. Nachr., 338, 249
\bibitem{} Reeves J.N. \et  \ 2003, ApJ, 593, 142
\bibitem{} Reeves J.N. \et  \ 2005, ApJ, 633, L81
\bibitem{} Reeves J.N. \et  \ 2018, ApJ, 854, 28R
\bibitem{} Shakura N.I. \& Sunyaev R.A. \ 1973, A\&A, 24, 337
\bibitem{} Strueder L. \et  \ 2001,  A\&A, 365, L18
\bibitem{} Tombesi F., Cappi M., Reeves J.N., Palumbo G.C., Yaqoob T., Braito V., Dadina M. \ 2010, ApJ, 742, 44
\bibitem{} Tombesi F., Cappi M., Reeves J.N., Palumbo G.C., Braito V., Dadina M. \ 2011, A\&A, 521, A57 
\bibitem{} Turner M.J.L. \et  \ 2001,  A\&A, 365, L27
\end{thebibliography}
\end{document}